\documentclass[aps,prl,twocolumn,showpacs,amsmath,amssymb,superscriptaddress]{revtex4} 

\usepackage{graphicx}  
\usepackage{epsfig}
\usepackage{dcolumn}   
\usepackage{bm}        
\usepackage{amssymb}   

\usepackage{color}
\usepackage[normalem]{ulem}
\definecolor{red}{rgb}{0.5,0,0}
\definecolor{blue}{rgb}{0,0,1}
\definecolor{green}{rgb}{0,0.5,0}
\definecolor{pink}{rgb}{0.5,0,0.5}
\definecolor{grey}{rgb}{0.7,0.7,0.7}

\def\mod2#1{\big|#1\big|^2}                 
\def\-{\!-\!}                               
\def\+{\!+\!}                               
\def\={\;=\;}                               

\def\x{\vec{x}}

\def\0{\vec{0}}

\def\d{\mathrm{d}}                          
\def\intd#1{\int\d#1\;}                     
\def\intd2#1{\int\d^2#1\;}                  
\def\intd3#1{\int\d^3#1\;}                  
\def\vec#1{\mathbf{#1}}                     
\def\-{\!-\!}                               
\def\+{\!+\!}                               

\def\beq{\begin{equation}}                  
\def\eeq{\end{equation}}                    

\begin{document}

\title{Quantitative chemical mapping at the atomic scale}

\author{Huolin L. Xin} \email{hx35@cornell.edu} \affiliation{Department of Physics, Cornell University, Ithaca, NY 14853, USA}
\author{Christian Dwyer} \email{christian.dwyer@monash.edu}\affiliation{Monash Centre for Electron Microscopy, ARC Centre of Excellence for Design in Light Metals, and Department of Materials Engineering, Monash University, Clayton 3800 Vic, Australia}
\author{David A. Muller} \email{dm24@cornell.edu} \affiliation{School of Applied and Engineering Physics and Kavli Institute at Cornell for Nanoscale Science, Cornell University, Ithaca, NY 14853, USA}


\begin{abstract}
Atomic-scale mapping of the chemical elements in materials is now possible using aberration-corrected electron microscopes but delocalization and multiple scattering can confound image interpretation. Here we report atomic-resolution measurements with the elastic and inelastic signals acquired on an absolute scale. By including dynamical scattering in both the elastic and inelastic channels we obtain quantitative agreement between theory and experiment. Our results enable a close scrutiny of the inelastic scattering physics and demonstrate the possibility of element-specific atom counting.

\end{abstract}

\maketitle


The short de Broglie wavelength of high-energy electrons enables a scanning transmission electron microscope (STEM) to form an electron beam of sub-atomic dimensions. Such beams can be used to probe the atomic and electronic structure of materials with extremely high selectivity. Collection of the inelastic scattering resulting from the excitation of atomic core electrons, via electron energy-loss spectroscopy (EELS), provides both chemical and electronic bonding information. Recently it has become possible to utilize the core-loss EELS signal in a STEM to map the chemical composition and electronic structure of materials at the atomic scale \cite{Okunishi2006, Bosman2007, Kimoto2007, Muller2008, Muller2009, Botton2010}. This capacity constitutes an extremely powerful tool in materials characterization. In the present work, we consider whether it is possible to perform \emph{quantitative} chemical mapping at the atomic scale. Such a capability would permit, for example, element-specific column-by-column atom counting, with significant implications for the materials and condensed matter communities. To this end, it is crucial to establish a sufficiently accurate interpretation of the contrast in atomic-resolution chemical maps, and, in particular, to determine whether the dominant inelastic scattering is treated correctly in current theories based on a single-particle description of the core-excitation process. Here we report the first atomic-resolution chemical mapping with the EELS signal acquired on an absolute scale (i.e.\ normalized with respect to the incident beam). We demonstrate that quantification within a single-particle picture is indeed possible under favorable circumstances. In other cases, we find discrepancies between the simulations and experiments indicative of a break-down of the single-particle model.

In a STEM, images are generated by scanning a convergent electron beam across an electron-transparent specimen in a raster-like fashion, while collecting various scattered signals in synchronization with the scanning beam. A common mode of atomic-resolution imaging in the STEM is annular dark-field (ADF) imaging, whereby the electrons elastically scattered to high angles are collected by an annular-shaped detector. While atomic-resolution ADF imaging of crystals can now be achieved routinely, quantification of the image contrast has a long and controversial history \cite{Stobbs1994,Howie2004}. Discrepancies between experimental and simulated contrast levels, referred to as the ``Stobbs factor'', were often found to be as large as 3--5 times, and led to speculation that fundamental scattering mechanisms were missing from the simulations. With improved quantification techniques, however, recent work \cite{LeBeau2008} has persuasively argued that discrepancies in ADF image contrast can be accommodated by including a physically-reasonable source size in the simulations. A major conclusion of that work is that multislice calculations \cite{GoodmanMoodie1974} incorporating a frozen-phonon model of thermal diffuse scattering \cite{Loane1991} correctly describe the dominant scattering mechanisms in ADF imaging. 

On the other hand, core-loss EELS offers a wealth of information not accessible by ADF imaging, including unique identification of the chemical elements and bonding information. However, the inelastic scattering processes relevant to core-loss EELS  add a level of complexity over and above that of ADF imaging. In this study, we have performed carefully calibrated experiments, allowing atomic-resolution chemical signals to be extracted on an absolute scale. By simultaneously recording the elastic signal, also on an absolute scale, and using it to characterize the electron beam and sample thickness, we eliminate all free parameters in the acquisition of the core-loss signals. Coupled with simulations that incorporate both core-loss inelastic scattering and dynamical elastic scattering, the present work enables a close scrutiny of the scattering physics in this inelastic channel.

Experiments were performed using an aberration-corrected Nion UltraSTEM operating at 100 kV with a beam convergence semi-angle of $31.8\pm0.1$ mrad. The specimen consisted of a DyScO$_3$ single crystal in the $[101]$ crystallographic orientation. The EELS signal was recorded using a Gatan Enfina spectrometer with a collection semi-angle of $79.8\pm0.1$ mrad. During all experiments, ADF images were recorded simultaneously with the energy-loss spectra. The inner and outer ADF collection semi-angles were $98.2\pm0.2$ and $294.5\pm0.5$ mrad, respectively. Both the EELS and ADF signals were recorded so as to allow their calibration in terms of the fractional beam intensity. Data was obtained for a large range of specimen thicknesses. This permits a direct comparison with theoretical predictions of absolute cross sections and chemical maps from the Dy-$M_{4,5}$ ($3d\rightarrow4f$), Dy-$N_{4,5}$ ($4d\rightarrow 4f$ and continuum), and Sc-$L_{2,3}$ ($2p\rightarrow3d$ and continuum) edges. The different symmetries of the Dy and Sc sublattices, and the presence of two different Dy edges, allow us to identify and separate elastic and inelastic effects. 

\begin{figure}[b]
\includegraphics[width=8.5cm]{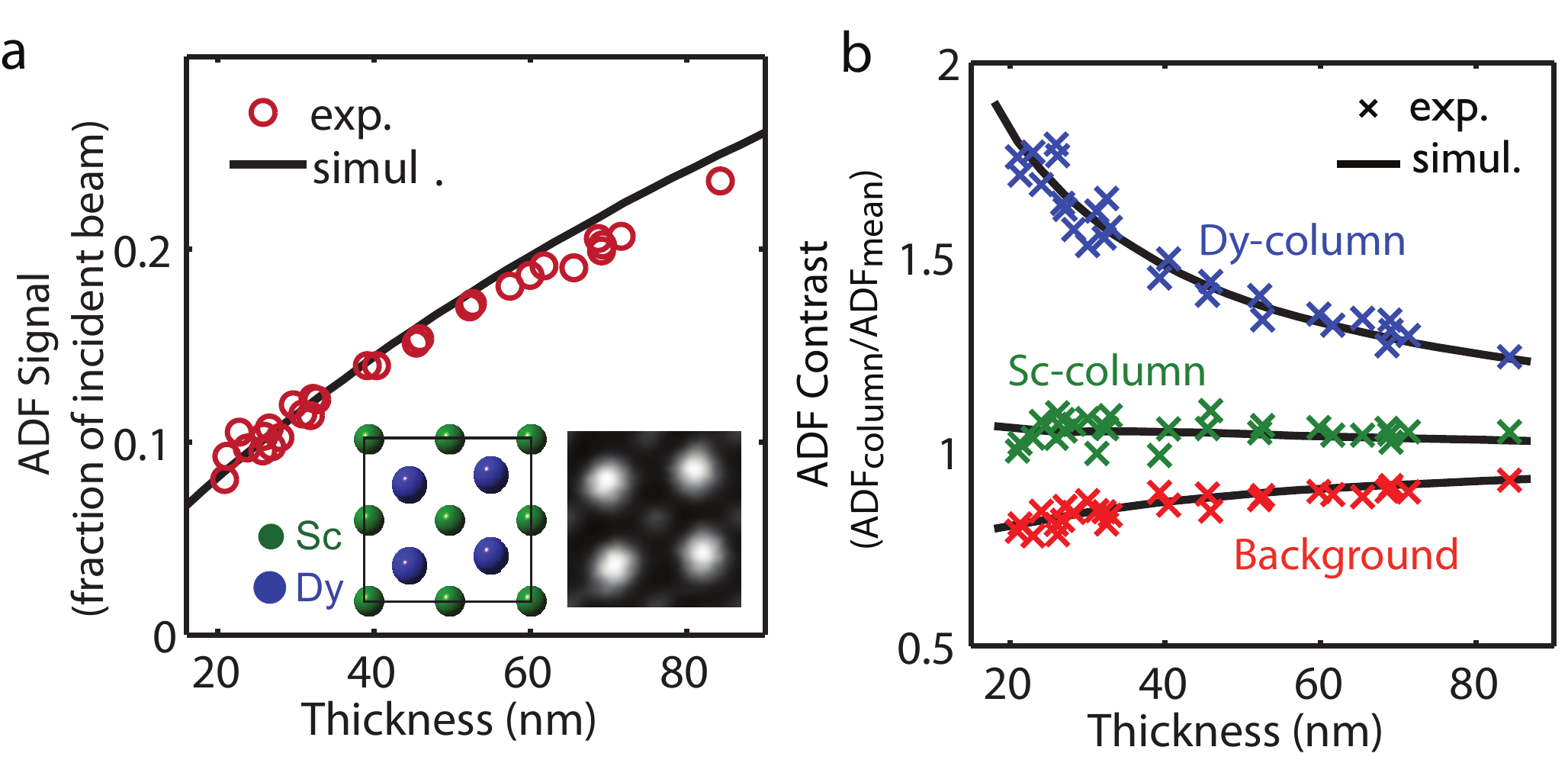}
\caption{Quantitative determination of the effective source distribution from the simultaneously-acquired ADF-STEM images. (a) Comparison of the mean ADF signal with the frozen phonon simulation on an absolute scale. (b) Comparison of the experimental image contrast with the frozen-phonon simulation incorporating the the determined source distribution.}
\label{fig:Quantitative_ADF}
\end{figure}

As a first step, we performed quantification of the ADF images. Not only does this provide a check on various experimental parameters, it also enables us to determine the effective source size of the instrument independently from the EELS measurements. Since source size effects are manifest in the same way in both ADF and EELS \cite{Dwyer2010_sourcesize}, this step is crucial for the quantitative analysis of the core-loss signal. ADF quantification is shown in Fig.~\ref{fig:Quantitative_ADF}. The inset of Fig.~\ref{fig:Quantitative_ADF}(a) shows an ADF image of one unit cell of $[101]$ DyScO$_3$. The ADF image consists of bright spots at the positions of Dy atomic columns and weaker spots at the positions of Sc atomic columns (O atoms are not discernible). The graph in Fig.~\ref{fig:Quantitative_ADF}(a) compares the experimental position-averaged, or mean, ADF signal as a function of sample thickness with the prediction from frozen phonon multislice simulations. The general agreement is very good (to within about 5\%). A quantitative comparison of the contrast in experimental and simulated images enables us to extract the effective source size of the instrument. Fig.~\ref{fig:Quantitative_ADF}(b) compares the experimental and simulated contrasts of Dy columns, Sc columns and the background. Here the contrast is defined as the ADF signal at the relevant position divided by mean ADF signal, eliminating any errors in the absolute cross sections. The effective source distribution was modeled as a combination of Gaussian and truncated-Lorentzian distributions, fitted using nonlinear least-square optimization. The effective source of our instrument is described by a Gaussian with a full-width at half-maximum (FWHM) of 0.71 \AA\ convolved with a truncated Lorentizian with a FWHM of 0.18 \AA\ that reflects residual aberrations in the instrument alignment.

The simulations used for quantitative analysis of the core-loss EELS signal were based on a relativistic multi-pole core-loss theory \citep{Dwyer2005, Dwyer2005_relativistic}, performed on a GPU cluster \cite{Dwyer2010_gpu}. The relevant scattering physics is described by the following expression for the wave function of an inelastically-scattered electron at the specimen exit surface:
\beq\label{eq:inelastic wave} \psi_\alpha(\x) = \int\d^2\x''\d^2\x'\, G_{\alpha}(\x,\x'')\sigma_\alpha V_{\alpha}(\x'')G_{0}(\x'',\x')\psi_0(\x'). \eeq
In this expression, $\psi_0$ is the wave function of the high-energy electron beam at the specimen entrance surface, $V_{\alpha}$ is a quantum-mechanical matrix element describing the effect of core excitation on the high-energy electron, $\sigma_\alpha$ is the interaction constant, and $G_{0}$ and $G_{\alpha}$ are the Green's functions describing the dynamical elastic scattering, or ``channeling'', of the high-energy electron before and after the core excitation event. Channeling was calculated using a frozen-phonon multislice method, while the matrix elements $V_{\alpha}$ were calculated within a single-particle model where the initial and final states of the core electron are described by atomic wave functions \cite{Dwyer2005, LeapmanRezMayers1980}. Since Eq.~\eqref{eq:inelastic wave} includes channeling before and after core excitation, it is referred to as a ``double-channeling'' theory \citep{RossouwMaslen1984, Allen-etal2006, Dwyer2008_doublechanneling}. We find that although a single-channeling approximation, obtained by neglecting channeling after core excitation \citep{Allen2003,Bosman2007,Wang2010_PRL}, produces qualitatively similar results, double-channeling effects are important for quantitative agreement, especially in the presence of heavy elements such as Dy.

\begin{figure*}[t]
\includegraphics[width=14.0cm]{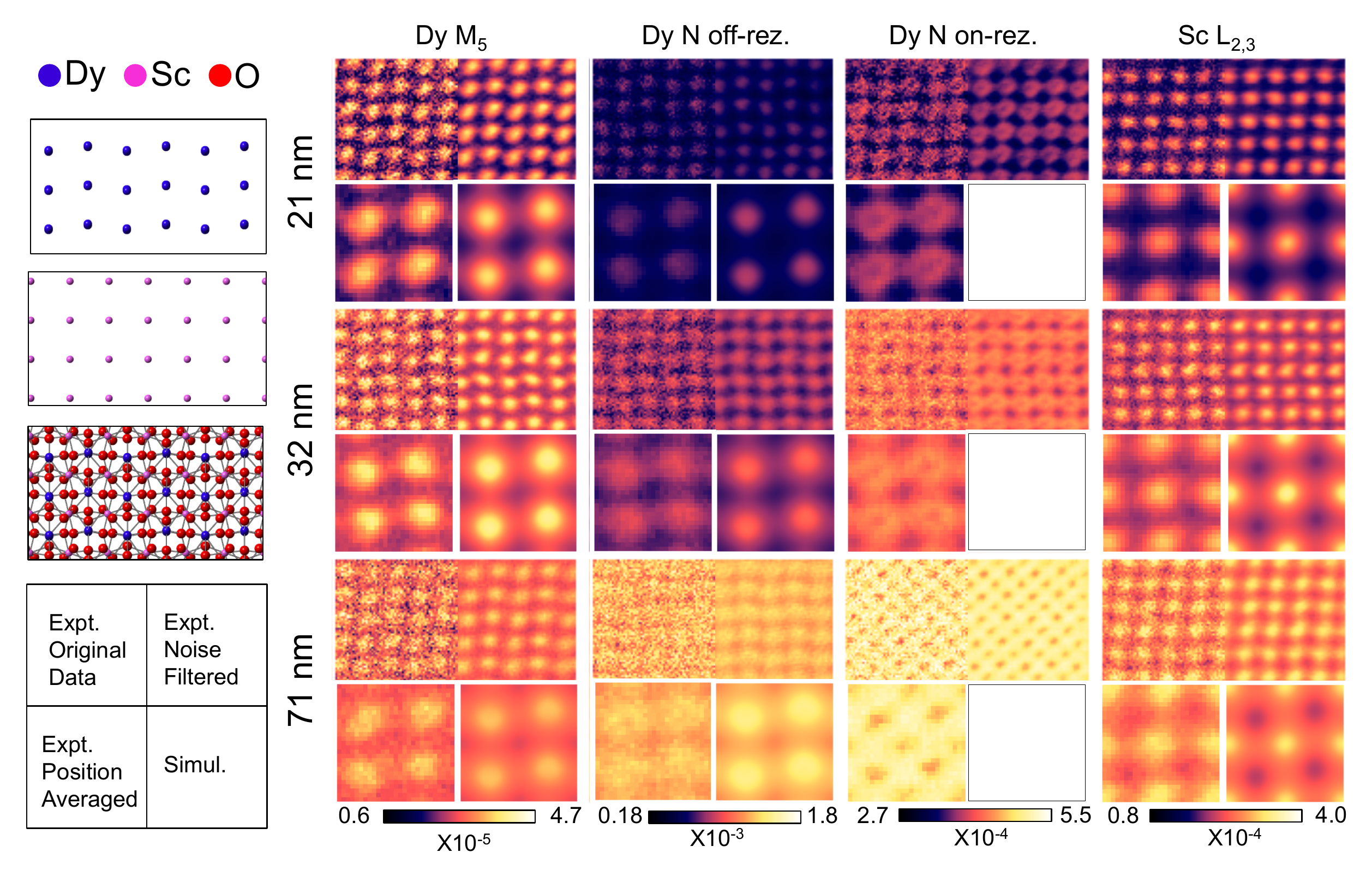}\caption{Quantitative atomic-resolution chemical mapping of $[101]$ DyScO$_3$. For each chemical signal (indicated atop) and specimen thickness (indicated left) we show raw, noise-filtered, unit-cell-averaged, and simulated maps. Unit-cell-averaged maps are from about 25 unit cells. Intensity scales refer to the fractional beam intensity. The experimental spectra and energy-loss integration windows are shown in Fig.~\ref{fig:Quantitative_EELS}(a). Simulations incorporate the overall scaling factors from Fig.~\ref{fig:Quantitative_EELS}(b).}
\label{fig:Quantitative_collage}
\end{figure*}

Fig.~\ref{fig:Quantitative_collage} shows a collage of experimental and simulated  chemical maps of $[101]$ DyScO$_3$ for a range of specimen thicknesses. The majority of the maps are seen to exhibit local maxima at the sites of the corresponding atomic columns for all thicknesses in the recorded range. This is a highly desirable feature, as it allows an intuitive interpretation of the maps in terms of the locations of atomic sites. These features are well-reproduced by the simulations that incorporate the effective source distribution. On the other hand, simulations that do not incorporate the effective source (not shown) reveal that in the Dy-$M_5$ and Dy-$N_{4,5}$ maps local minima do, in fact, develop at the Dy columns as the thickness increases beyond about 20 nm. While similar minima have been observed previously and were attributable to coherent effects \cite{Kohl1985, Muller1995, Allen2003_atomicresEELS, Cosgriff2005_volcanoes, Bosman2007, Dwyer2008_doublechanneling, DAlfonso2008_volcanoes}, our large collection angle implies that here the minima develop as a consequence of strong elastic scattering by the heavy Dy column prior to core excitation, with the scattering subsequent to core excitation also playing a role. For the Dy-$N_{4,5}$ maps, the form of the matrix elements is such that the minima are even more pronounced, to the point where the Dy-$N_{4,5}$ maps in Fig.~\ref{fig:Quantitative_collage} can exhibit minima at the Dy columns even when the effective source is included. The effect is particularly strong in the Dy-$N_{4,5}$ on-resonance map, which exhibits clear minima across almost the entire range of thicknesses. Contrary to this behavior, the Sc-$L_{2,3}$ maps exhibit no such minima because the lighter Sc columns imply weaker elastic scattering. Overall, the results in Fig.~\ref{fig:Quantitative_collage} demonstrate very good agreement between the simulated and experimental chemical maps across the range of recorded thicknesses. 

\begin{figure}
\includegraphics[width=9.5cm]{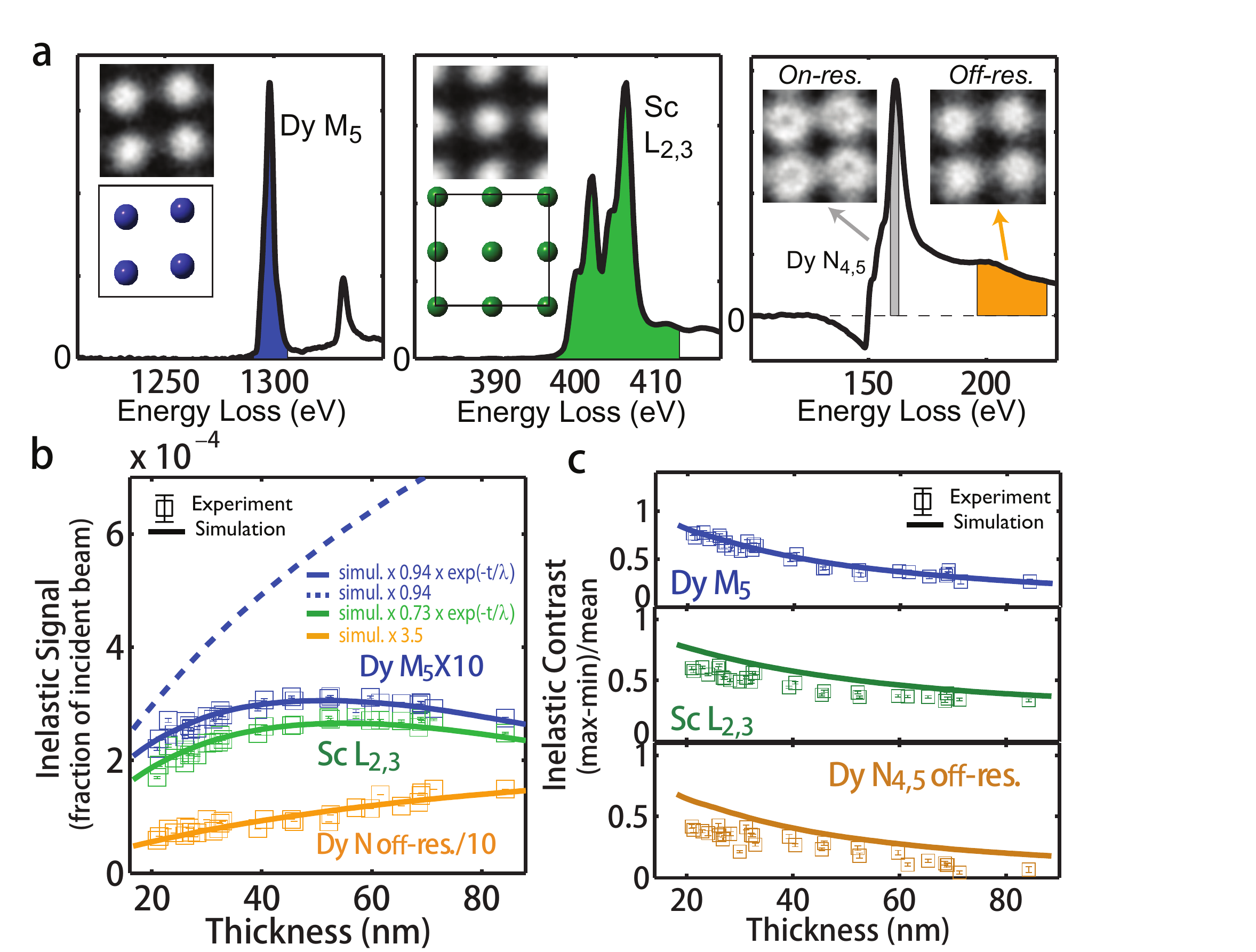}
\caption{Quantitative comparison of the inelastic signal and chemical mapping contrast. (a) Background-subtracted experimental spectra at the Dy-$M_5$, Sc-$L_{2,3}$ and Dy-$N_{4,5}$ edges. (b) Comparison of experimental and simulated mean inelastic signals. (c) Comparison of experimental and simulated chemical mapping contrast.}
\label{fig:Quantitative_EELS}
\end{figure}

Let us proceed to a quantitative comparison of the experimental and simulated chemical maps. The energy-loss spectra and integration windows of the signals are shown in Fig.~\ref{fig:Quantitative_EELS}(a). For Dy-$M_5$ and Sc-$L_{2,3}$ we have analyzed the near-edge signals since they are strongest, while for Dy-$N_{4,5}$ we have avoided complications associated with the Fano resonance and have analyzed the signal at 60 eV above threshold. We proceed by comparing the absolute cross-sections, as represented by the mean inelastic signals as a function of sample thickness, as shown in Fig.~\ref{fig:Quantitative_EELS}(b). Firstly, it is noted that for Dy-$M_5$, the simulation significantly overestimates the mean signal (dashed line in Fig.~\ref{fig:Quantitative_EELS}(b)). The discrepancy is attributable to inelastic scattering associated with multiple plasmon excitations, which is absent from the simulations. Ideally, the effects of such scattering would be removed from the experimental spectra by deconvolution. However, we find that this method is unreliable in the case of atomic-resolution EELS where the low signal means that the spectra are noisy and the extended edges are not fully recorded. Instead, we chose to incorporate the effects of multiple plasmon scattering into the simulations as a correction given by Beer's law $e^{-t/\lambda_0}$, where $t$ is the sample thickness and $\lambda_0$ is the inelastic mean-free path. This correction has been applied to the simulated Dy-$M_5$ and Sc-$L_{2,3}$ signals in Fig.~\ref{fig:Quantitative_EELS}(b). After applying the correction, we observe that the thickness-dependent trends of the simulated mean signals agree well with the experiments. Moreover, the absolute value of the Dy-$M_5$ mean signal is in agreement with experiment to within about 5\%. For the Sc-$L_{2,3}$ signal, we observe that the simulation overestimates the signal by about 37\%. For the Dy-$N_{4,5}$ off-resonance signal, no Beer's law correction was applied because the integration window receives signal from the resonance while losing signal at the same time. The simulation overestimates the Dy-$N_{4,5}$ signal by a factor slightly less than 3. As discussed below, the larger discrepancies for Sc-$L_{2,3}$ and Dy-$N_{4,5}$ indicate a break-down of either the single-particle model or assumptions in the processing of the experimental data. Overall, the mean inelastic signals are in agreement with the simulations within the expected uncertainties, but highlight the need to include multiple inelastic scattering. 

Similar trends of agreement carry over to the contrasts in the chemical maps, as presented in Fig.~\ref{fig:Quantitative_EELS}(c). For example, the Dy-$M_5$ contrast is reproduced closely by the simulation (when the latter includes the effective source distribution). The Sc-$L_{2,3}$ contrast is slightly overestimated by the simulation, the discrepancy being around 25\% for thicknesses below 50 nm, and reducing to about 15\% for thicknesses greater than 50 nm. For the Dy-$N_{4,5}$ off-resonance maps, the discrepancies are more significant, with the experimental contrast being lower by nearly a factor of 2. Nonetheless, the overall thickness-dependent trends of the experimental and simulated contrasts are in agreement for both Sc-$L_{2,3}$ and Dy-$N_{4,5}$.

The discrepancies observed for Sc-$L_{2,3}$ are likely due to one of three factors. Firstly, the $L_{2,3}$ branching-ratio anomaly of Sc indicates that there is a large core hole--3$d$ interaction, implying that the single-particle model used in the simulations is not fully correct. Secondly, the final states for the Sc-$L_{2,3}$ signal are sensitive to solid state effects not included in the simulations. In line with these observations, comparison of the experimental and simulated Sc-$L_{2,3}$ mean signal at 60 eV above threshold (not shown) exhibits substantially better agreement. Thirdly, the Sc-$L_{2,3}$ edge is sitting on a background signal whose spatial distribution changes with energy loss. In the processing of experimental data, an extrapolation of the pre-edge intensities is used to extract the core-loss signals. The assumption of this procedure is that the extrapolated background has the same spatial distribution as the pre-edge signal, and this assumption can be invalid at low energy losses. In the case of Dy-$N_{4,5}$, the observed discrepancies are likely caused by the strong Fano resonance \citep{Fano1961,Dietz1974,Rez2001}. As shown in Fig.~\ref{fig:Quantitative_EELS}(a), the Dy-$N_{4,5}$ pre-edge intensities are negative due to the Fano resonance, making the background subtraction problematic. In addition, as exemplified in Fig.~\ref{fig:Quantitative_collage}, the signal in the Dy-$N_{4,5}$ on-resonance maps spreads throughout the unit cell due to the relatively large spatial extent of the 4$d$ and 4$f$ orbitals of Dy. Even though the Dy-$N_{4,5}$ off-resonance maps were extracted from about 50 eV above the resonance, it is possible that a significant fraction of the signal originates from the resonance. In contrast to the Sc-$L_{2,3}$ and Dy-$N_{4,5}$ signals, the situation for Dy-$M_5$ is relatively simple. For example, the atomic-like final states for the Dy-$M_5$ signal fully justify the single-particle atomic model used to compute the quantum-mechanical matrix elements. Moreover, the relatively large energy loss gives rise to a slowly-varying background signal, ensuring validity of the background subtraction.

In conclusion, we have demonstrated the simultaneous acquisition of quantitative ADF images and EELS maps in an aberration-corrected STEM. Coupled with double-channeling simulations of core-loss inelastic scattering, this has enabled a quantitative examination of the core-loss signal utilized in atomic-resolution chemical mapping. We found that by taking into account the effective source distribution determined from the ADF images, both the absolute signal and the contrast in atomic-resolution Dy-$M_5$ maps can be closely reproduced by the double-channeling simulations. At lower energy losses, discrepancies are present in the Sc-$L_{2,3}$ and Dy-$N_{4,5}$ maps due to the energy-dependent spatial distribution of the background spectrum, core-hole effects, and omitted complexities in the final states. This work has demonstrated the possibility of using quantitative STEM-EELS for element-specific column-by-column atom counting at higher energy losses and for atomic-like final states, and has elucidated several possible improvements for future work.

Funded by the Energy Materials Center at Cornell, an Energy Frontier Research Center (DOE $\#$DE-SC0001086). C.D. would like to thank Philip Chan, eSolutions-Research Support Services and the Monash e-Research Centre for the use of Monash Central HPC cluster, and acknowledges financial support from the Australian Research Council (DP110104734).


\begin{thebibliography}{29}
\expandafter\ifx\csname natexlab\endcsname\relax\def\natexlab#1{#1}\fi
\expandafter\ifx\csname bibnamefont\endcsname\relax
  \def\bibnamefont#1{#1}\fi
\expandafter\ifx\csname bibfnamefont\endcsname\relax
  \def\bibfnamefont#1{#1}\fi
\expandafter\ifx\csname citenamefont\endcsname\relax
  \def\citenamefont#1{#1}\fi
\expandafter\ifx\csname url\endcsname\relax
  \def\url#1{\texttt{#1}}\fi
\expandafter\ifx\csname urlprefix\endcsname\relax\def\urlprefix{URL }\fi
\providecommand{\bibinfo}[2]{#2}
\providecommand{\eprint}[2][]{\url{#2}}

\bibitem[{\citenamefont{Okunishi et~al.}(2006)\citenamefont{Okunishi, Sawada,
  Kondo, and Kersker}}]{Okunishi2006}
\bibinfo{author}{\bibfnamefont{E.}~\bibnamefont{Okunishi}},
  \bibinfo{author}{\bibfnamefont{H.}~\bibnamefont{Sawada}},
  \bibinfo{author}{\bibfnamefont{Y.}~\bibnamefont{Kondo}}, \bibnamefont{and}
  \bibinfo{author}{\bibfnamefont{M.}~\bibnamefont{Kersker}},
  \bibinfo{journal}{Microsc. Microanal.} \textbf{\bibinfo{volume}{12 (Supp
  2)}}, \bibinfo{pages}{1150} (\bibinfo{year}{2006}).

\bibitem[{\citenamefont{Bosman et~al.}(2007)\citenamefont{Bosman, Keast,
  Garc\'\i{}a-Mu\~noz, D'Alfonso, Findlay, and Allen}}]{Bosman2007}
\bibinfo{author}{\bibfnamefont{M.}~\bibnamefont{Bosman}},
  \bibinfo{author}{\bibfnamefont{V.~J.} \bibnamefont{Keast}},
  \bibinfo{author}{\bibfnamefont{J.~L.} \bibnamefont{Garc\'\i{}a-Mu\~noz}},
  \bibinfo{author}{\bibfnamefont{A.~J.} \bibnamefont{D'Alfonso}},
  \bibinfo{author}{\bibfnamefont{S.~D.} \bibnamefont{Findlay}},
  \bibnamefont{and} \bibinfo{author}{\bibfnamefont{L.~J.} \bibnamefont{Allen}},
  \bibinfo{journal}{Phys. Rev. Lett.} \textbf{\bibinfo{volume}{99}},
  \bibinfo{pages}{086102} (\bibinfo{year}{2007}).

\bibitem[{\citenamefont{Kimoto et~al.}(2007)\citenamefont{Kimoto, Asaka, Nagai,
  Saito, Matsui, and Ishizuka}}]{Kimoto2007}
\bibinfo{author}{\bibfnamefont{K.}~\bibnamefont{Kimoto}},
  \bibinfo{author}{\bibfnamefont{T.}~\bibnamefont{Asaka}},
  \bibinfo{author}{\bibfnamefont{T.}~\bibnamefont{Nagai}},
  \bibinfo{author}{\bibfnamefont{M.}~\bibnamefont{Saito}},
  \bibinfo{author}{\bibfnamefont{Y.}~\bibnamefont{Matsui}}, \bibnamefont{and}
  \bibinfo{author}{\bibfnamefont{K.}~\bibnamefont{Ishizuka}},
  \bibinfo{journal}{Nature} \textbf{\bibinfo{volume}{450}},
  \bibinfo{pages}{702} (\bibinfo{year}{2007}).

\bibitem[{\citenamefont{Muller et~al.}(2008)\citenamefont{Muller, Kourkoutis,
  Murfitt, Song, Hwang, Silcox, Dellby, and Krivanek}}]{Muller2008}
\bibinfo{author}{\bibfnamefont{D.~A.} \bibnamefont{Muller}},
  \bibinfo{author}{\bibfnamefont{L.~F.} \bibnamefont{Kourkoutis}},
  \bibinfo{author}{\bibfnamefont{M.}~\bibnamefont{Murfitt}},
  \bibinfo{author}{\bibfnamefont{J.~H.} \bibnamefont{Song}},
  \bibinfo{author}{\bibfnamefont{H.~Y.} \bibnamefont{Hwang}},
  \bibinfo{author}{\bibfnamefont{J.}~\bibnamefont{Silcox}},
  \bibinfo{author}{\bibfnamefont{N.}~\bibnamefont{Dellby}}, \bibnamefont{and}
  \bibinfo{author}{\bibfnamefont{O.~L.} \bibnamefont{Krivanek}},
  \bibinfo{journal}{Science} \textbf{\bibinfo{volume}{319}},
  \bibinfo{pages}{1073} (\bibinfo{year}{2008}).

\bibitem[{\citenamefont{Muller}(2009)}]{Muller2009}
\bibinfo{author}{\bibfnamefont{D.~A.} \bibnamefont{Muller}},
  \bibinfo{journal}{Nature Materials} \textbf{\bibinfo{volume}{8}},
  \bibinfo{pages}{263} (\bibinfo{year}{2009}).

\bibitem[{\citenamefont{Botton et~al.}(2010)\citenamefont{Botton, Lazar, and
  Dwyer}}]{Botton2010}
\bibinfo{author}{\bibfnamefont{G.~A.} \bibnamefont{Botton}},
  \bibinfo{author}{\bibfnamefont{S.}~\bibnamefont{Lazar}}, \bibnamefont{and}
  \bibinfo{author}{\bibfnamefont{C.}~\bibnamefont{Dwyer}},
  \bibinfo{journal}{Ultramicroscopy} \textbf{\bibinfo{volume}{110}},
  \bibinfo{pages}{926 } (\bibinfo{year}{2010}).

\bibitem[{\citenamefont{Hytch and Stobbs}(1994)}]{Stobbs1994}
\bibinfo{author}{\bibfnamefont{M.}~\bibnamefont{Hytch}} \bibnamefont{and}
  \bibinfo{author}{\bibfnamefont{W.}~\bibnamefont{Stobbs}},
  \bibinfo{journal}{Ultramicroscopy} \textbf{\bibinfo{volume}{53}},
  \bibinfo{pages}{191} (\bibinfo{year}{1994}).

\bibitem[{\citenamefont{Howie}(2004)}]{Howie2004}
\bibinfo{author}{\bibfnamefont{A.}~\bibnamefont{Howie}},
  \bibinfo{journal}{Ultramicroscopy} \textbf{\bibinfo{volume}{98}},
  \bibinfo{pages}{73} (\bibinfo{year}{2004}).

\bibitem[{\citenamefont{LeBeau et~al.}(2008)\citenamefont{LeBeau, Findlay,
  Allen, and Stemmer}}]{LeBeau2008}
\bibinfo{author}{\bibfnamefont{J.~M.} \bibnamefont{LeBeau}},
  \bibinfo{author}{\bibfnamefont{S.~D.} \bibnamefont{Findlay}},
  \bibinfo{author}{\bibfnamefont{L.~J.} \bibnamefont{Allen}}, \bibnamefont{and}
  \bibinfo{author}{\bibfnamefont{S.}~\bibnamefont{Stemmer}},
  \bibinfo{journal}{Phys. Rev. Lett.} \textbf{\bibinfo{volume}{100}}
  (\bibinfo{year}{2008}).

\bibitem[{\citenamefont{Goodman and Moodie}(1974)}]{GoodmanMoodie1974}
\bibinfo{author}{\bibfnamefont{P.}~\bibnamefont{Goodman}} \bibnamefont{and}
  \bibinfo{author}{\bibfnamefont{A.~F.} \bibnamefont{Moodie}},
  \bibinfo{journal}{Acta Cryst. A} \textbf{\bibinfo{volume}{30}},
  \bibinfo{pages}{280} (\bibinfo{year}{1974}).

\bibitem[{\citenamefont{Loane et~al.}(1991)\citenamefont{Loane, Xu, and
  Silcox}}]{Loane1991}
\bibinfo{author}{\bibfnamefont{R.~F.} \bibnamefont{Loane}},
  \bibinfo{author}{\bibfnamefont{P.}~\bibnamefont{Xu}}, \bibnamefont{and}
  \bibinfo{author}{\bibfnamefont{J.}~\bibnamefont{Silcox}},
  \bibinfo{journal}{Acta Cryst. A} \textbf{\bibinfo{volume}{47}},
  \bibinfo{pages}{267} (\bibinfo{year}{1991}).

\bibitem[{\citenamefont{Dwyer et~al.}(2010)\citenamefont{Dwyer, Erni, and
  Etheridge}}]{Dwyer2010_sourcesize}
\bibinfo{author}{\bibfnamefont{C.}~\bibnamefont{Dwyer}},
  \bibinfo{author}{\bibfnamefont{R.}~\bibnamefont{Erni}}, \bibnamefont{and}
  \bibinfo{author}{\bibfnamefont{J.}~\bibnamefont{Etheridge}},
  \bibinfo{journal}{Ultramicroscopy} \textbf{\bibinfo{volume}{110}},
  \bibinfo{pages}{952} (\bibinfo{year}{2010}).

\bibitem[{\citenamefont{Dwyer}(2005{\natexlab{a}})}]{Dwyer2005}
\bibinfo{author}{\bibfnamefont{C.}~\bibnamefont{Dwyer}},
  \bibinfo{journal}{Ultramicroscopy} \textbf{\bibinfo{volume}{104}},
  \bibinfo{pages}{141} (\bibinfo{year}{2005}{\natexlab{a}}).

\bibitem[{\citenamefont{Dwyer}(2005{\natexlab{b}})}]{Dwyer2005_relativistic}
\bibinfo{author}{\bibfnamefont{C.}~\bibnamefont{Dwyer}},
  \bibinfo{journal}{Phys. Rev. B} \textbf{\bibinfo{volume}{72}}
  (\bibinfo{year}{2005}{\natexlab{b}}).

\bibitem[{\citenamefont{Dwyer}(2010)}]{Dwyer2010_gpu}
\bibinfo{author}{\bibfnamefont{C.}~\bibnamefont{Dwyer}},
  \bibinfo{journal}{Ultramicroscopy} \textbf{\bibinfo{volume}{110}},
  \bibinfo{pages}{195} (\bibinfo{year}{2010}).

\bibitem[{\citenamefont{Leapman et~al.}(1980)\citenamefont{Leapman, Rez, and
  Mayers}}]{LeapmanRezMayers1980}
\bibinfo{author}{\bibfnamefont{R.~D.} \bibnamefont{Leapman}},
  \bibinfo{author}{\bibfnamefont{P.}~\bibnamefont{Rez}}, \bibnamefont{and}
  \bibinfo{author}{\bibfnamefont{D.~F.} \bibnamefont{Mayers}},
  \bibinfo{journal}{J. Chem. Phys.} \textbf{\bibinfo{volume}{72}},
  \bibinfo{pages}{1232} (\bibinfo{year}{1980}).

\bibitem[{\citenamefont{Rossouw and Maslen}(1984)}]{RossouwMaslen1984}
\bibinfo{author}{\bibfnamefont{C.~J.} \bibnamefont{Rossouw}} \bibnamefont{and}
  \bibinfo{author}{\bibfnamefont{V.~M.} \bibnamefont{Maslen}},
  \bibinfo{journal}{Phil. Mag. A} \textbf{\bibinfo{volume}{49}},
  \bibinfo{pages}{743} (\bibinfo{year}{1984}).

\bibitem[{\citenamefont{Allen et~al.}(2006)\citenamefont{Allen, Findlay, Oxley,
  Witte, and Zaluzec}}]{Allen-etal2006}
\bibinfo{author}{\bibfnamefont{L.~J.} \bibnamefont{Allen}},
  \bibinfo{author}{\bibfnamefont{S.~D.} \bibnamefont{Findlay}},
  \bibinfo{author}{\bibfnamefont{M.~P.} \bibnamefont{Oxley}},
  \bibinfo{author}{\bibfnamefont{C.}~\bibnamefont{Witte}}, \bibnamefont{and}
  \bibinfo{author}{\bibfnamefont{N.~J.} \bibnamefont{Zaluzec}},
  \bibinfo{journal}{Phys. Rev. B} \textbf{\bibinfo{volume}{73}},
  \bibinfo{pages}{094104} (\bibinfo{year}{2006}).

\bibitem[{\citenamefont{Dwyer et~al.}(2008)\citenamefont{Dwyer, Findlay, and
  Allen}}]{Dwyer2008_doublechanneling}
\bibinfo{author}{\bibfnamefont{C.}~\bibnamefont{Dwyer}},
  \bibinfo{author}{\bibfnamefont{S.~D.} \bibnamefont{Findlay}},
  \bibnamefont{and} \bibinfo{author}{\bibfnamefont{L.~J.} \bibnamefont{Allen}},
  \bibinfo{journal}{Phys. Rev. B} \textbf{\bibinfo{volume}{77}}
  (\bibinfo{year}{2008}).

\bibitem[{\citenamefont{Allen et~al.}(2003{\natexlab{a}})\citenamefont{Allen,
  Findlay, Oxley, and Rossouw}}]{Allen2003}
\bibinfo{author}{\bibfnamefont{L.~J.} \bibnamefont{Allen}},
  \bibinfo{author}{\bibfnamefont{S.~D.} \bibnamefont{Findlay}},
  \bibinfo{author}{\bibfnamefont{M.~P.} \bibnamefont{Oxley}}, \bibnamefont{and}
  \bibinfo{author}{\bibfnamefont{C.~J.} \bibnamefont{Rossouw}},
  \bibinfo{journal}{Ultramicroscopy} \textbf{\bibinfo{volume}{96}},
  \bibinfo{pages}{47} (\bibinfo{year}{2003}{\natexlab{a}}).

\bibitem[{\citenamefont{Wang et~al.}(2010)\citenamefont{Wang, Behan, Takeguchi,
  Hashimoto, Mitsuishi, Shimojo, Kirkland, and Nellist}}]{Wang2010_PRL}
\bibinfo{author}{\bibfnamefont{P.}~\bibnamefont{Wang}},
  \bibinfo{author}{\bibfnamefont{G.}~\bibnamefont{Behan}},
  \bibinfo{author}{\bibfnamefont{M.}~\bibnamefont{Takeguchi}},
  \bibinfo{author}{\bibfnamefont{A.}~\bibnamefont{Hashimoto}},
  \bibinfo{author}{\bibfnamefont{K.}~\bibnamefont{Mitsuishi}},
  \bibinfo{author}{\bibfnamefont{M.}~\bibnamefont{Shimojo}},
  \bibinfo{author}{\bibfnamefont{A.~I.} \bibnamefont{Kirkland}},
  \bibnamefont{and} \bibinfo{author}{\bibfnamefont{P.~D.}
  \bibnamefont{Nellist}}, \bibinfo{journal}{Phys. Rev. Lett.}
  \textbf{\bibinfo{volume}{104}}, \bibinfo{pages}{200801}
  (\bibinfo{year}{2010}).

\bibitem[{\citenamefont{Kohl and Rose}(1985)}]{Kohl1985}
\bibinfo{author}{\bibfnamefont{H.}~\bibnamefont{Kohl}} \bibnamefont{and}
  \bibinfo{author}{\bibfnamefont{H.}~\bibnamefont{Rose}},
  \bibinfo{journal}{Adv. Imag. Electr. Phys.} \textbf{\bibinfo{volume}{65}},
  \bibinfo{pages}{173} (\bibinfo{year}{1985}).

\bibitem[{\citenamefont{Muller and Silcox}(1995)}]{Muller1995}
\bibinfo{author}{\bibfnamefont{D.~A.} \bibnamefont{Muller}} \bibnamefont{and}
  \bibinfo{author}{\bibfnamefont{J.}~\bibnamefont{Silcox}},
  \bibinfo{journal}{Ultramicroscopy} \textbf{\bibinfo{volume}{59}},
  \bibinfo{pages}{195} (\bibinfo{year}{1995}).

\bibitem[{\citenamefont{Allen et~al.}(2003{\natexlab{b}})\citenamefont{Allen,
  Findlay, Lupini, Oxley, and Pennycook}}]{Allen2003_atomicresEELS}
\bibinfo{author}{\bibfnamefont{L.~J.} \bibnamefont{Allen}},
  \bibinfo{author}{\bibfnamefont{S.~D.} \bibnamefont{Findlay}},
  \bibinfo{author}{\bibfnamefont{A.~R.} \bibnamefont{Lupini}},
  \bibinfo{author}{\bibfnamefont{M.~P.} \bibnamefont{Oxley}}, \bibnamefont{and}
  \bibinfo{author}{\bibfnamefont{S.~J.} \bibnamefont{Pennycook}},
  \bibinfo{journal}{Phys. Rev. Lett.} \textbf{\bibinfo{volume}{91}},
  \bibinfo{pages}{105503} (\bibinfo{year}{2003}{\natexlab{b}}).

\bibitem[{\citenamefont{Cosgriff et~al.}(2005)\citenamefont{Cosgriff, Oxley,
  Allen, and Pennycook}}]{Cosgriff2005_volcanoes}
\bibinfo{author}{\bibfnamefont{E.~C.} \bibnamefont{Cosgriff}},
  \bibinfo{author}{\bibfnamefont{M.~P.} \bibnamefont{Oxley}},
  \bibinfo{author}{\bibfnamefont{L.~J.} \bibnamefont{Allen}}, \bibnamefont{and}
  \bibinfo{author}{\bibfnamefont{S.~J.} \bibnamefont{Pennycook}},
  \bibinfo{journal}{Ultramicroscopy} \textbf{\bibinfo{volume}{102}},
  \bibinfo{pages}{317} (\bibinfo{year}{2005}).

\bibitem[{\citenamefont{{D'Alfonso} et~al.}(2008)\citenamefont{{D'Alfonso},
  Findlay, Oxley, and Allen}}]{DAlfonso2008_volcanoes}
\bibinfo{author}{\bibfnamefont{A.~J.} \bibnamefont{{D'Alfonso}}},
  \bibinfo{author}{\bibfnamefont{S.~D.} \bibnamefont{Findlay}},
  \bibinfo{author}{\bibfnamefont{M.~P.} \bibnamefont{Oxley}}, \bibnamefont{and}
  \bibinfo{author}{\bibfnamefont{L.~J.} \bibnamefont{Allen}},
  \bibinfo{journal}{Ultramicroscopy} \textbf{\bibinfo{volume}{108}},
  \bibinfo{pages}{677} (\bibinfo{year}{2008}).

\bibitem[{\citenamefont{Fano}(1961)}]{Fano1961}
\bibinfo{author}{\bibfnamefont{U.}~\bibnamefont{Fano}}, \bibinfo{journal}{Phys.
  Rev.} \textbf{\bibinfo{volume}{124}}, \bibinfo{pages}{1866}
  (\bibinfo{year}{1961}).

\bibitem[{\citenamefont{Dietz et~al.}(1974)\citenamefont{Dietz, McRae, Yafet,
  and Caldwell}}]{Dietz1974}
\bibinfo{author}{\bibfnamefont{R.~E.} \bibnamefont{Dietz}},
  \bibinfo{author}{\bibfnamefont{E.~G.} \bibnamefont{McRae}},
  \bibinfo{author}{\bibfnamefont{Y.}~\bibnamefont{Yafet}}, \bibnamefont{and}
  \bibinfo{author}{\bibfnamefont{C.~W.} \bibnamefont{Caldwell}},
  \bibinfo{journal}{Phys. Rev. Lett.} \textbf{\bibinfo{volume}{33}},
  \bibinfo{pages}{1372} (\bibinfo{year}{1974}).

\bibitem[{\citenamefont{Rez}(2001)}]{Rez2001}
\bibinfo{author}{\bibfnamefont{P.}~\bibnamefont{Rez}},
  \bibinfo{journal}{Microsc. Microanal.} \textbf{\bibinfo{volume}{7}},
  \bibinfo{pages}{356} (\bibinfo{year}{2001}).

\end{thebibliography}
\end{document}